\documentclass[twocolumn,aps,pra,floatfix,superscriptaddress]{revtex4-2}
\usepackage[braket, qm]{qcircuit}
\usepackage{mathptmx}
\usepackage{amsmath}
\usepackage{amssymb}
\usepackage{braket}
\usepackage{graphicx}
\usepackage{amsfonts}
\usepackage{csquotes}
\usepackage{hhline}
\usepackage{relsize}
\usepackage{listings}
\usepackage{color}
\usepackage{qcircuit}
\usepackage{physics}
\usepackage{soul} 
\usepackage{dsfont}

\usepackage{adjustbox}

\usepackage{float}
\usepackage{tikz}
\usetikzlibrary{positioning}
\lstdefinestyle{mystyle}{
    backgroundcolor=\color{backcolour},   
    commentstyle=\color{codegreen},
    keywordstyle=\color{magenta},
    numberstyle=\tiny\color{codegray},
    stringstyle=\color{codepurple},
    basicstyle=\footnotesize,
    breakatwhitespace=false,         
    breaklines=true,                 
    captionpos=b,                    
    keepspaces=true,                 
    numbers=left,                    
    numbersep=5pt,                  
    showspaces=false,                
    showstringspaces=false,
    showtabs=false,                  
    tabsize=2
}
 
\usepackage{subfigure}

\usepackage{dcolumn}
\usepackage{tabularx}
\usepackage[colorlinks=true,linkcolor=blue,citecolor=blue,urlcolor=blue]{hyperref}

\newcommand{\iu}{\text{i}}
\newcommand{\eu}{\text{e}}

\usepackage{dcolumn}% Align table columns on decimal point
\usepackage{cases}
\usepackage{multirow}
\usepackage{xcolor}
\makeatletter
\let\newfloat\newfloat@ltx
\makeatother

\usepackage{algorithm,algorithmic}

\begin{document}

\setstcolor{red}

\title{Coherent Feed Forward Quantum Neural Network} 
\author{Utkarsh Singh}
\affiliation{National Research Council of Canada, 100 Sussex Drive, Ottawa, Ontario K1N 5A2, Canada}
\affiliation{Department of Physics, University of Ottawa, 25 Templeton Street, Ottawa, Ontario, K1N 6N5 Canada}
\author{Aaron Z. Goldberg}
\affiliation{National Research Council of Canada, 100 Sussex Drive, Ottawa, Ontario K1N 5A2, Canada}
 \affiliation{Department of Physics, University of Ottawa, 25 Templeton Street, Ottawa, Ontario, K1N 6N5 Canada}
\author{Khabat Heshami}
\affiliation{National Research Council of Canada, 100 Sussex Drive, Ottawa, Ontario K1N 5A2, Canada}
\affiliation{Department of Physics, University of Ottawa, 25 Templeton Street, Ottawa, Ontario, K1N 6N5 Canada}
 \affiliation{Institute for Quantum Science and Technology, Department of Physics and Astronomy, University of Calgary, Alberta T2N 1N4, Canada}

%\date{July 2023}

\begin{abstract}
%Machine learning tasks can be improved using quantum devices
%Conventional computers may be outstripped by quantum devices performing machine learning tasks
%Quantum devices could use entanglement to process data in large Hilbert spaces and outperform conventional computers at machine learning tasks.
Quantum machine learning, focusing on quantum neural networks (QNNs), remains a vastly uncharted field of study. Current QNN models primarily employ variational circuits on an ansatz or a quantum feature map, often requiring multiple entanglement layers. This methodology not only increases the computational cost of the circuit beyond what is practical on near-term quantum devices but also misleadingly labels these models as neural networks, given their divergence from the structure of a typical feed-forward neural network (FFNN). Moreover, the circuit depth and qubit needs of these models scale poorly with the number of data features, resulting in an efficiency challenge for real-world machine-learning tasks. We introduce a \textit{bona fide} QNN model, which seamlessly aligns with the versatility of a traditional FFNN in terms of its adaptable intermediate layers and nodes, absent from intermediate measurements such that our entire model is coherent. This model stands out with its reduced circuit depth and number of requisite C-NOT gates to outperform prevailing QNN models. Furthermore, the qubit count in our model remains unaffected by the data's feature quantity. We test our proposed model on various benchmarking datasets such as the diagnostic breast cancer (Wisconsin) and credit card fraud detection datasets. We compare the outcomes of our model with the existing QNN methods to showcase the advantageous efficacy of our approach, even with a reduced requirement on quantum resources. Our model paves the way for application of quantum neural networks to real relevant machine learning problems.
\end{abstract}

\maketitle

%\tableofcontents

\section{Introduction \label{n_Sw1}}

Over the past decade, quantum machine learning (QML)~\cite{Schuld} has emerged as a dynamic field with promising potential for advancing machine learning techniques using quantum computing, especially because quantum machines may efficiently explore large-dimensional spaces for processing large amounts of data~\cite{NielsonChuang, qml1, qml2, qml5, qml16, qml15}.  While initial research focused primarily on adapting standard machine learning algorithms to quantum computing, such as quantum neural networks (QNNs) \cite{qml6, qml9, qml11, qml12, qml13, qml14} and quantum support vector machines \cite{qml16, qml10}, progress has been hindered by the lack of a clear path to scalability and practical applications of these methods. Instead, researchers have focused on developing algorithms suitable for the current generation of noisy intermediate-scale quantum (NISQ) devices~\cite{Wan2017Sep,Beer2020Feb, Bondarenko2020Mar, Cong2019Dec, qml14, Sharma2022May, Zhou2023May}, resulting in a new wave of algorithms known as NISQ-era QML algorithms~\cite{Preskill2018Aug}.

These recent QML algorithms are predicated upon low depth parameterized quantum circuits~\cite{Sim2019Dec, Zhang2023Feb}, which take a hybrid approach that combines the strengths of quantum processors with classical processors. This hybridization allows for the development of novel algorithms that have shown some advantages in specific use cases~\cite{qml1, Jiang2021Jan, Yamasaki2023Dec}, although a useful quantum advantage remains to be seen. 

Most QML models currently available employ complex encoding techniques, known as quantum feature maps, and use parameterized quantum circuits as models \cite{qml15, qml16} in place of the intermediate layers of a neural network. Sometimes, these circuits are concatenated such that the measurement of one circuit provides nonlinearity when inputting data into the subsequent circuit~\cite{Tacchino2020Oct}. In the hybrid approach, post-processing steps are similar to those used in classical machine learning (ML), updating the parametrized quantum circuits using techniques such as gradient descent, and are performed on a classical computer. Nonetheless, these feature mapping techniques often stumble when faced with real-world datasets, as the requisite number of qubits and the circuit depth escalate with the number of features in the data. %\st{We will introduce these current QML models in more detail in section} \ref{sec22} \aaron{(\st{I am not sure that this sentence is required, unless we also have sentences elsewhere saying "section blank will talk about blank, etc" but I don't think the nature style needs this})}. 
% In addition, the term "quantum neural network" is somewhat misleading for these models, since they do not adjust intermediate layers and neurons as a general neural network would. 

In this work, we propose a novel circuit-based approach that incorporates entanglement layers for the connections between nodes in adjacent layers, resembling a classical feed-forward neural network (FFNN). This approach offers the advantage of adaptable intermediate layers, allowing us to adjust them according to the characteristics of the data, which is particularly significant for classification tasks. Further, because all of the hidden layers can be incorporated without intermediate measurements, our approach is fully coherent throughout the evolution of the circuit and can take advantage of quantum coherence properties throughout, which is responsible for quantum advantages in related settings \cite{Gyurik2023arxiv}. Finally, by developing a data encoding scheme inspired by classical neural networks that writes multiple features onto a small number of qubits, our overall use of quantum resources is amenable to quantum computers available today. %\aaron{\st{Finally, by developing a data encoding scheme that writes multiple features onto small number of qubits, our overall use of quantum resources is amenable to quantum computers available today.}} 

To evaluate the effectiveness of our model, we conduct numerical experiments using the credit card fraud detection and Wisconsin breast cancer diagnostic datasets, employing the Qiskit package for simulations of quantum circuits, and compare the results to state-of-the-art QNN models. The results of these experiments are presented show that our approach achieves significant improvements in both accuracy and computational efficiency over traditional QNN methods. 

Our results highlight the potential of integrating classical neural network concepts into quantum computing frameworks,
opening avenues for more sophisticated, resource-efficient quantum models in the future. As we continue to
explore and refine our model that we dub the coherent feed-forward quantum neural network (CFFQNN), we anticipate its adaptability to a broader range of applications, further bridging the gap between quantum computing and real-world machine learning challenges.

% \section{Background}
\subsection{Artificial Neural Network \label{sec21}}
% \khabat{This subsection is too long. We should consider cutting the unnecessary parts or what is mostly common knowledge.} \aaron{I shortened a bunch, have listed one paragraph that is a candidate for deletion, more condensing might still be needed.}

%The functioning of the brain inspires the idea of an artificial neural network (ANN). While a complete understanding of the internal mechanisms of the brain is elusive we know that the brain receives and process information via a network of neurons, also known as axon-synapse-dendrite connections. \khabat{the next couple of sentences should be passive} We use a similar structure to process the data and the information in a neural network. We call them artificial neurons or perceptrons or nodes. These nodes are connected to other nodes via links similar to those inside the brain. Each link has a weight, which determines the strength of one node's influence on another. A perceptron represents a simplified model of a biological neuron, designed to mimic its basic function: receiving inputs, processing them, and producing an output.An artificial neuron is shown in figure \ref{fig:sub1}. \khabat{Fig. 2 is referred to before Fig. 1!}

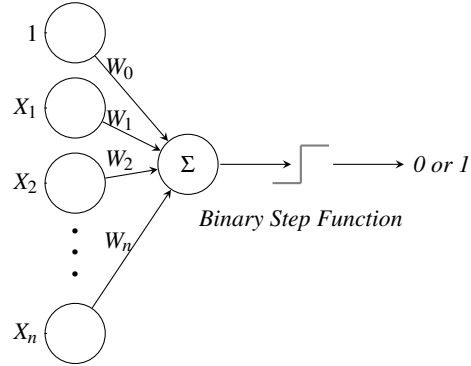
\begin{figure}[h!]
\centering

\def\layersep{1cm}
\tikzset{%
  every neuron/.style={
    circle,
    draw,
    text height=0.333cm,
    minimum size=0.8cm
  },
  neuron missing/.style={
    draw=none, 
    scale=2,
    text height=0.233cm,
    execute at begin node=\color{black}$\vdots$
  },
    neuron empty/.style={
    draw=none, 
    scale=3,
    text height=0.233cm
  },
}

\begin{tikzpicture}[x=0.75cm, y=1cm, >=stealth]

\foreach \m/\l [count=\y] in { 1,2,3,missing,4}
  \node [every neuron/.try, neuron \m/.try] (input-\m) at (0,2-\y) {};

\foreach \m [count=\y] in {1}
  \node [ every neuron/.try, neuron \m/.try ] (hidden1-\m) at (2,-0.75) {};
  
\foreach \m [count=\y] in {empty}
  \node [ every neuron/.try, neuron \m/.try ] (hidden11) at (2.2,-0.75) {};
  
  \foreach \m [count=\y] in {empty}
  \node [every neuron/.try, neuron \m/.try ] (hidden2) at (4.2,-0.75) {};

\foreach \m [count=\y] in {empty}
  \node [every neuron/.try, neuron \m/.try ] (output) at (6.2,-0.75) {};

% naming 
\foreach \l [count=\i] in {1,X_{1},X_2,X_n}
  \draw [] (input-\i) -- ++(-0.5,0)
    node [left] {$\l$};

\foreach \l [count=\i] in {W_{0}}
  \node [font = {\small}] at (0.8,0.52) {$\l$};
\foreach \l [count=\i] in {W_{1}}
  \node [font = {\small}] at (0.8,-0.16) {$\l$};
\foreach \l [count=\i] in {W_{2}}
  \node [font = {\small}] at (0.8,-0.7) {$\l$};
\foreach \l [count=\i] in {W_{n}}
  \node [font = {\small}] at (0.78,-1.78) {$\l$};
    
\foreach \l [count=\i] in {\Sigma}
  \node [font = {\small}] at (2,-0.75) {$\l$};
  
\foreach \l [count=\i] in {\textit{Binary Step Function}}
  \node [font = {\small}] at (4,-1.5) {$\l$};
  
\draw[gray, thick] (4,-0.5) -- (4.5,-0.5);
\draw[gray, thick] (4,-0.5) -- (4,-1);
\draw[gray, thick] (3.5,-1) -- (4,-1);

% \foreach \l [count=\i] in {\mathbf{_{-}\mid^{-}}}
%   \node [font = {\small}] at (4, -0.75) {$\l$};

\foreach \l [count=\i] in {\textit{0 or 1}}
  \node [font = {\small}] at (6.5, -0.75) {$\l$};
  
% connections 
\foreach \i in {1,...,4}
  \foreach \j in {1}
    \draw [->] (input-\i) -- (hidden1-\j);
    
\foreach \i in {1}
    \draw [<-] (hidden11) -- (hidden2);

\foreach \j in {1}
    \draw [<-] (hidden2) -- (output);

% \foreach \l [count=\x from 0] in {Input, 1st Hidden, 2nd Hidden, Ouput}
%   \node [align=center, above] at (\x*2,2) {\l \\ layer};
  
\end{tikzpicture}

  \caption{A perceptron, inspired by neural networks in the brain. Inputs $\{X_i\}$ are combined with weights $\{W_i\}$ including a bias $W_0$ that are processed nonlinearly to produce a binary output.}
  \label{fig:sub1}
\end{figure}

The functioning of the brain inspires the idea of an artificial neural network (ANN). The brain receives and processes information via a network of neurons, where each neuron receives inputs from a number of neurons, processes them, and produces an output that is then input to subsequent neurons. In an ANN, perceptrons or nodes are used to mimic biological neurons: each is linked to others with variable weights and the structure of the connections between nodes and their weights can then be used to process data;
one artificial neuron is shown in Fig.~\ref{fig:sub1}. 
%\khabat{\st{Fig. 2 is referred to before Fig. 1!}}

Perceptrons were introduced by Frank Rosenblatt in 1957~\cite{rosenblatt1957perceptron} as binary classifiers that form the foundational concept behind artificial neural networks and deep learning. A perceptron takes multiple input values $X_i$ and produces a single binary-outcome output $y$. Each input is associated with a weight $W_i$, which signifies the importance of that input. The perceptron computes a weighted sum of its inputs with an overall bias $b$ via 
\begin{equation}
    z=\sum_i W_i X_i+b
\end{equation} and passes this sum through an activation function, typically a step function, to produce its output $y$:
\begin{equation}
    y=\sigma(z).
\end{equation} Here, to make it a differentiable non-linear activiation function, $\sigma $ is typically taken to be a logistic sigmoid or rectified linear unit function.
If the weighted sum exceeds a certain threshold, the perceptron outputs a 1 (or ``active''), otherwise, it outputs a 0 (or ``inactive'').

The perceptron's strength lies in its simplicity, adept at modelling linearly separable data, but it falters with non-linear data. This spurred the evolution of multi-layer perceptrons (MLPs) or neural networks capable of handling complex, non-linear patterns. ANNs optimize weights in each layer using methods like backpropagation and gradient descent, allowing them to approximate any function with high accuracy~\cite{Cybenko1989Dec,Hornik1991Jan}. FFNNs, a key type of ANN, allow unidirectional information flow and have shown impressive performance in tasks like classification and regression. Any exemplary FFNN will be shown below in Fig.~\ref{fig5}(a).

\subsection{Quantum Neural Network \label{sec22}}    

\begin{figure*}[hbt!]
\centering
\includegraphics[width=\textwidth]{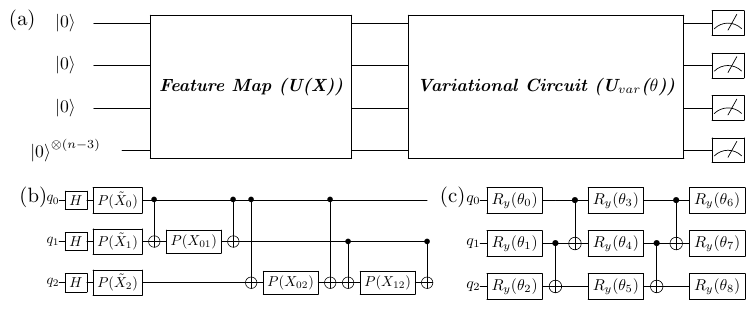}
\caption{(a) Architecture of a QNN acting on $n$ qubits; the data $\mathbf{X}$ are loaded with a feature map $U(\mathbf{X})$ and the data are processed using a parametrized circuit $U_{var}(\theta)$. Subsequent measurement allows the parameters $\theta$ to be optimized and updated. These steps may be repeated. (b) A 3-qubit feature map circuit administering the commonly used ZZFeatureMap. Here $H$ represents the Hadamard gate, $P$ represents the phase gate, $\tilde{X}_{i} = 2X_{i}$, and $X_{ij} = 2(\pi - X_{i})(\pi - X_{j})$. (c) A 3-qubit variational circuit with weight parameters $\{\theta_j\}$ explicit.}
  \label{fig1}
\end{figure*}

Quantum neural networks are among the most popular algorithms in QML. Even though the field is not fully developed, there is a rapidly growing set of people exploring potential quantum advantages~\cite{cancers15102705, Bondarenko2020Mar, Jiang2021Jan, Farhi2018}. 

The momentum behind recent advancements in this domain can largely be attributed to the variational techniques prevalent in numerous hybrid algorithms \cite{qml15}. In general, one starts by encoding the classical data $\mathbf{X} \in \mathbb{R}^{N}$ onto the quantum state using a feature map $U(\mathbf{X})$ that can be implemented by a quantum circuit~\cite{qiskit-textbook}. Next, a variational circuit $U_{var}(\boldsymbol{\theta})$ is applied with intelligently chosen single-qubit rotations $R(\theta_i)$ and entanglement layers on an input state $U(\mathbf{X})\ket{0}^{\otimes n}$, where the $\theta_i$ parameters are the trainable weights and $n$ is the number of qubits employed and usually scales linearly with $N$. Finally, the expectation value of some observable (e.g. $\hat{Z}^{\otimes n}$) is measured for the classical post-processing to predict the class $\Tilde{y}$ of the input data. 
Optimizing the weight parameters present in the variational circuit is done using some classical optimizers, eventually minimizing the cost function $C(  y(\mathbf{X},\boldsymbol{\theta}),\Tilde{y})$
in consideration~\cite{qml15}. A schematic diagram of a typical QNN is shown in figure \ref{fig1}.

While this approach of QNNs demonstrates promising outcomes in certain applications, it faces substantial challenges in scalability and gate complexity, particularly on NISQ devices~\cite{Cerezo2022Sep, Beer2020Feb}. The number of qubits required in QNNs tends to increase linearly with the number of data features, quickly exceeding the limited qubit capacity of current quantum hardware and thus restricting their applicability to smaller datasets, instead of making use of the exponential scaling of Hilbert space dimension with number of qubits. Additionally, the number of controlled-NOT (C-NOT) gates, crucial for creating entanglements in quantum circuits, scales nearly quadratically with the number of features. This scaling is problematic on NISQ devices due to increased error rates and circuit depth, leading to higher likelihoods of decoherence and computational inefficiency. The combination of these scalability issues with the gate complexity challenges significantly hinders the practical implementation of QNNs on existing quantum platforms, making the handling of complex, feature-rich datasets a formidable task and posing a significant bottleneck in fully leveraging quantum computing for advanced machine learning applications.
 
\section{Results \label{res}}
\subsection{The Model: Coherent Feed-Forward Quantum Neural Network \label{mthd}}

\begin{figure}[H]
    \centering
\includegraphics[width=0.49\textwidth]{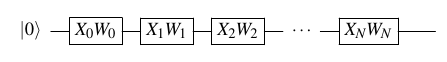}
\caption{ The depiction of data encoding stage. Rotation gates with angles $X_0W_0$ act on a single qubit in analogy with inputs acting on a single neuron. An extra rotation gate with $X_0=1$ and $W_0=b$ is added for flexibility to bias the initial qubit.}
\label{fig4}
\end{figure}

\begin{figure}[H]
    \centering
    \includegraphics[width=0.25\textwidth]{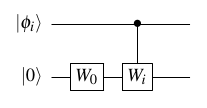}

\caption{An illustration of a single intermediate node in CFFQNN and how the weight values are applied from one node to the next using controlled rotations by angle $W_i$, with a possible single-qubit rotation by bias angle $W_0$.}
\label{intermediate}
\end{figure}

\begin{figure*}[hbt!]
\centering
  \includegraphics*[viewport=0 9 365 368, scale=1.05]{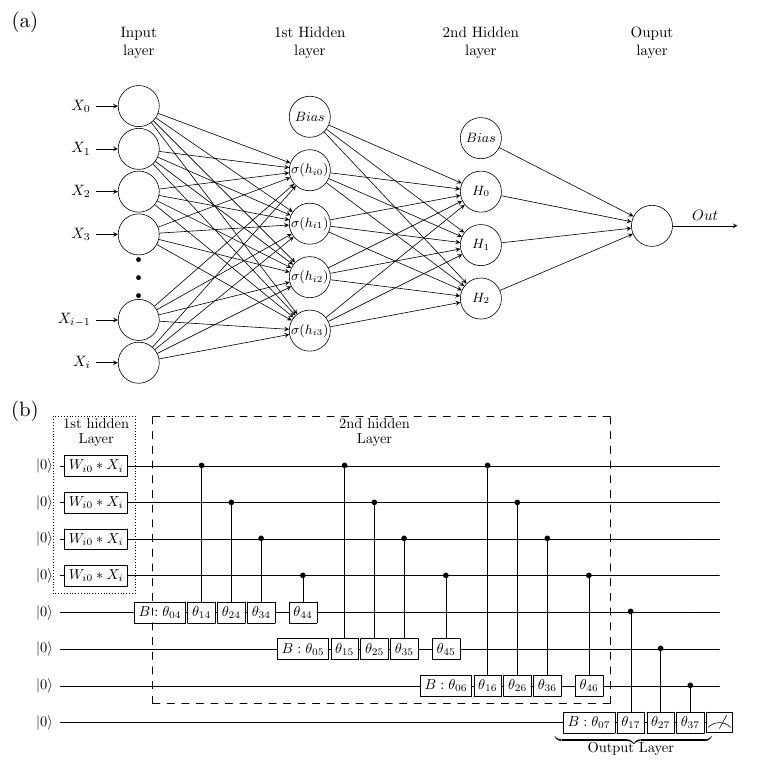}
  \caption{(a) Architecture of an artificial neural network with two hidden layers. Here $W$ represents the weight parameters, $\mathbf{X}$ are data points, $\sigma$ is a non-linear activation function, and $h_{ij}=W_{ij}X_{i}$. (b) Architecture of a CFFQNN with two hidden layers where $\mathbf{X}$ are data points, and $\mathbf{W}$ and $\boldsymbol{\theta}$ represent the weight parameters. The number of modes in a layer of the ANN correspond to the number of qubits in a layer of the CFFQNN. In contrast to earlier QNN models such as those in Fig.~\ref{fig1}(c), the parameters of the CFFQNN change the controlled operations such that the CFFQNN circuits are not solely parametrized by their single-qubit gates; this is what allows the CFFQNN to resemble an ANN.}
  \label{fig5}
\end{figure*}

Here, we introduce our CFFQNN model that uses a quantum-classical hybrid approach to process data. The source code for all of our work can be found on GitHub as detailed below. %It processes classical data using a quantum-classical hybrid approach by encoding the classical data in a quantum system, manipulating the data using classically parametrized quantum circuits, measuring the quantum state, then performing classical computations to update the circuit parameters based on the measurement results. 
The initial encoding layer is similar to the first hidden layer of a conventional ANN, as illustrated in Fig.~\ref{fig4}, with the classical data loaded onto quantum states. Subsequent layers consist of a network single-qubit and controlled (entangling) rotation gates, all of which are parametrized by their rotation angles, exemplified in Fig.~\ref{intermediate}. Ultimately, the qubits undergo measurement. Notably, the parameterized controlled rotations are adaptable to cater to specific network demands and the measurement process can be tailored based on both the data and the desired structural outcome. This circuitry inherently mirrors the architecture of an ANN, as can be seen by comparing Fig.~\ref{fig5}(a) versus (b).

We elect to perform all rotations about the $y$-axis for reasons that will become clear shortly. These are mathematically described by the single-qubit rotation gate expressed in the single-qubit computational basis $\{|0\rangle,|1\rangle\}$ as
\begin{equation*}
    R_{y} (\theta)= \exp(-\iu\theta\sigma_y/2)=\begin{pmatrix}
    \cos \frac{\theta}{2} & -\sin \frac{\theta}{2} \\
    \sin \frac{\theta}{2} & \cos \frac{\theta}{2}
    \end{pmatrix}
\end{equation*} using the Pauli matrix $\sigma_y$. Sequential rotations about the same axis commute and act additively as $R_y(\theta_1)R_y(\theta_2)=R_y(\theta_1+\theta_2)$.

In the first layer, the data points with weights are encoded as the rotation angle of the $R_y$ gate acting on some initial state, where the latter is taken to be some general state $R_y(b)|0\rangle$. All of the data points are successively encoded onto the same qubit, which is schematized in Fig.~\ref{fig4}. Since all of the rotations are about the same axis, such an encoding can be performed with a single single-qubit gate parametrized by $\theta=z=\sum_{i=1}^N X_i W_i+b$:
\begin{equation}
    R_y(X_N W_N)\cdots R_y(X_2 W_2)R_y(X_1 W_1)[R_y(b)|0\rangle]=R_y(z)|0\rangle.
\end{equation}
This is the first efficiency resulting from all of the rotations being about the same axis, which reduces the number of data-encoding gates, and also is responsible for better mimicking an ANN by directly encoding the variable $z$ without giving a preference to the \textit{ordering} among nodes within a given layer.

%\[
%\Qcircuit @C=1em @R=3em {
%& \lstick{\ket{0}} & \gate{X_{0}W_{0}}& \gate{X_{1}W_{1}} & \gate{X_{2}W_{2}} &  \qw &   \stick{\cdots} &  &  \gate{X_{N}W_{N}} &  \qw & \qw %\\ }
%\]

This data encoding procedure can be repeated on multiple distinct qubits to allow for more nonlinear processing of the data in the quantum circuit. Therefore, a CFFQNN with $n$ qubits in the first layer can be described as 
\begin{equation}
    \begin{split}
       % \ket{\Psi} &= U(z)_{1} \otimes U(z)_{2} \otimes \cdots \otimes U(z)_{n} \ket{0}^{\otimes n}
       \ket{\Psi} &= [R(z)|0\rangle]^{\otimes n}
    \end{split}.
    \label{eqn3}
\end{equation} 
Such single gates are exemplified in the ``1st hidden layer'' segment of Fig.~\ref{fig5}(b).

Since $n$ is, in principle, independent from $N$, the size of the circuit need not to depend on the number of data features, unlike the case for a traditional QNN.

In the subsequent layers corresponding to hidden layers of an ANN, we initialize new qubits by applying a parametrized rotation gate as a biasing term $R_{y;\,j}(\theta_{0j})$ acting on the $j$th qubit, then we apply parametrized controlled rotation gates that are controlled by the qubits in the first layer. 
For example, if we want to connect the first node in the first layer to the first node in the second layer, we apply a rotation on the $(n+1)$th qubit controlled by the state of the $1$st qubit:
\begin{equation}
    CR_y^{1\mapsto n+1}(\theta)=|0\rangle_1\langle 0|\otimes \mathbb{I}_{n+1}+|1\rangle_1\langle 1|\otimes R_{y;\,n+1}(\theta).
\end{equation} 
The other nodes are similarly connected by controlled-rotation gates $CR_y^{i\mapsto j}(\theta_{ij})$, each parametrized by independent weights $\theta_{ij}$.
Even though the rotations are controlled by different control qubits in different states, all of the rotations on a new qubit commute, such that the aggregate effect on a node in a hidden layer is independent from any ordering among the nodes in the previous layer. 

%\[
%\Qcircuit @C=1em @R=2em {
%& \lstick{\ket{\phi_{i}}} & \qw & \ctrl{1} & \qw &  \qw &   \\ 
%& \lstick{\ket{0}} & \gate{W_{0}}& \gate{W_{i}} & \qw &  \qw &  \\ }
%\]

Figure \ref{intermediate} displays a single intermediate node in the CFFQNN's hidden layer. It illustrates how the node applies weight values (denoted as $W$) to these nodes depending on the value of the last layer nodes. After this operation, the state $\ket{\phi_{i}}\otimes\ket{0}$ changes to a new state $\ket{\psi_{ij}}=CR_{y}^{i\mapsto j}(\theta_{ij}) \ket{\phi_{i}}\otimes\ket{0}$ that may continue to be acted upon by other control layers.

While a classical ANN uses a perceptron to decide whether or not a certain node should forward its output to the next one, the CFFQNN retains the branches of the wavefunction associated with the control qubit being in each of states $|0\rangle$ and $|1\rangle$ without collapsing the state via measurement. For example, the branch of the state where all of the control qubits are in state $|0\rangle$ only applies the bias rotation to the target qubits, while the branch where all of the control qubits are in state $|1\rangle$ applies the gate $R_{y;\,j}(\sum_{0=1}^n\theta_{ij})$ to qubit $j$. Overall, the action takes the form
\begin{equation}
    U=\prod_{ij}CR_y^{i\mapsto j}(\theta_{ij})=\sum_{\mathbf{X}}|\mathbf{X}\rangle\langle \mathbf{X}|\otimes \bigotimes_j R_{y;\,j}\left(\sum_{i=0}^n X_i \theta_{ij}\right).
\end{equation}
Here, $\mathbf{X}={X_1,\cdots,X_n}$ is a bit string with elements $X_i\in \{0,1\}$, the sum runs over all such strings, and we have included $X_0=1$ to represent the bias term; the tensor product over $j$ implies that the rotation on the $j$th qubit associated with the branch of the wavefunction where the qubits are in state $|\mathbf{x}\rangle$ is by an angle $\sum_{i=0}^n X_i \theta_{ij}$, corresponding to the standard factor in ANNs. This structure is schematized in the ``2nd hidden layer'' segment of Fig.~\ref{fig5}(b), where the connections between control and target qubits are explicit and their weights are given accordingly. All output values of the perceptron are essentially kept coherently and the system is ready to have the process repeated in a subsequent layer.

% For example, operation on a hidden layer, other than the first layer, \st{qubit} can be written as

% \begin{equation*}
%     \begin{split}
%         &\ket{\psi} = (I \otimes W_{bias,j}) \times \prod CW_{ij} \ket{\phi_{i} 0}\\
%         & = (IR_{y}(\theta_{bias,j})) \times \prod CR_{y}(\theta_{ij}) \ket{\phi_{i} 0},
%     \end{split}
% \end{equation*} \label{eqn33}
% where the variables $i$ and $j$ respectively denote the indices of the control layer (the previous layer) and the current layer, i.e., $W_{i,j}$ represents the operation of $W$ on the $j_{th}$ layer depending on the values of $i_{th}$ layer. $W_{bias,j}$ represents the biasing term on $j_{th}$ layer. $R_{y}$ and $CR_{y}$ are the $Y$-rotation and Controlled $Y$-rotation gates, respectively. 

Put another way, we have replaced the nonlinear activation function $\sigma$ in a standard perceptron by the controlled operations $CR_y$. Unlike $\sigma$, the output of $CR_y$ is not deterministic; it is probabilistic. Nevertheless, if we measure one of the control qubits, we know we will find state $|0\rangle$ with probability $p(0)=\cos^2\frac{z}{2}$ times and state $|1\rangle$ with the rest of the probability $p(1)=\sin^2\frac{z}{2}$, depending on the value of $z=\sum_{i=1}^n X_i W_i+b$. We can turn such probabilities into binary outputs by simply choosing the larger of the two, which are split at the value $\alpha=\pi/2$:

\begin{equation}
     \Tilde{y}=\begin{cases}1,&
    p(1) > p(0)\iff   z > \alpha \\
    0, & p(1) < p(0)\iff  z <  \alpha
    \end{cases}.
\end{equation} In this sense, we have created a coherent QNN that retains all of the properties of an ANN while allowing for data to be processed without each perceptron being restricted to a binary output.

%We start the next layer by applying one biasing term (parameterized gate) on the qubit(s) and then other parameterized gates controlled by the qubits in the first layer, i.e., if the qubit in the first layer is in state 1, it will apply a parameterized $Ry$ gate on the next layer qubit(s).  
After repeating the process for multiple layers, the number of controlled (entangling) gates is given by the number of connections between the layers. This is at most a quadratic function of the number of nodes per layer and a linear function in the depth of the neural network, which are expected to grow with the number of features in the data but do not follow a fixed relationship. One can thus create a few-qubit quantum neural network with $n\ll N$ and verify empirically its success for a given machine-learning tasks.

Finally, after all of the intermediate (hidden) layers, we perform a measurement and calculate the expectation value of some operator, which we further use to decide the class of the data point as in a classical NN. In our work we often measure the value of the final ($N$th) qubit $\langle Z_{N}\rangle$ as depicted in the ``output layer'' segment of Fig.~\ref{fig5}(b). The measurement result is fed into a classical nonlinear activation function $\sigma$ to produce the binary outcome
\begin{equation}
    \Tilde{y} = \begin{cases}
    1 ,&   \sigma(\langle Z_N \rangle ) = 1 \\
    0 ,&  \sigma(\langle Z_N \rangle) = 0
    \end{cases}.
    \label{eq:perceptron from Z expectation}
\end{equation} More general outcomes can be considered by either dividing the ranges of expectation values $\langle Z_N\rangle$ into more than two segments or by measuring more final qubits to yield more possible final outcomes. The results of such outcomes can be used to update the encoding weights $W_i$ and intermediate weights $\theta_{ij}$ throughout the network in an iterative fashion.

This method can create a complete FFNN like a standard classical one without doing intermediate measurements. An overall schematic diagram of a CFFQNN with two hidden layers with four and three qubits respectively is shown in the figure \ref{fig5}(b).

\subsubsection{\textit{Different Variations of the Model}}

In our work, we deploy two distinct versions of the CFFQNN model: the standard CFFQNN and a variant that we dub FixedCFFQNN. While FixedCFFQNN retains the architectural design of the CFFQNN, it diverges in one key aspect: the weights in its initial layer remain untrained. This reduction in the number of parameters to be trained significantly speeds the training process while still outperforming previous QNNs.

\subsubsection{\textit{Hyperparameters of the CFFQNN Model}}

The architecture of CFFQNN closely mirrors that of an ANN, sharing many of the same hyperparameters. This allows for customization in terms of the number of layers and nodes within each layer. Additionally, the measurement scheme can be tailored to fit specific data needs; even the parameter $\alpha$ in Eq.~\eqref{eq:perceptron from Z expectation} is a hyperparameter. For instance, throughout our research, we employed a single measurement strategy for CFFQNN, measuring only the final qubit. For the FixedCFFQNN, we adopted a partial measurement approach, targeting all qubits except the qubits in the initial layer.

\subsection{Numerical Results}

 We evaluate the CFFQNN's performance against the prevailing QNN model across various datasets. For the conventional QNN model, we employ the ZZFeatureMap \cite{Qiskit} to encode classical data into the quantum circuit and the RealAmplitudes \cite{Qiskit} circuit as the  variational circuit with the trainable weights. We use the COBYLA \cite{scikit-learn} optimizer to optimize the weights ($\theta$s). Figure \ref{fig1} illustrates a 3-qubit ZZFeatureMap circuit alongside a RealAmplitudes circuit with two repetition layers.
Additionally, to provide a comprehensive performance perspective, we compared the efficacy of all quantum models against the classical MLPClassifier—an FFNN.

To predict the input data's output class, we predominantly employed the Statevector simulator from Qiskit \cite{Qiskit}\cite{qiskit-textbook}, designed to emulate the ideal quantum states of a quantum system without any external noise or decoherence. This simulator offers an exact depiction of the quantum state, facilitating precise calculations and forecasts. It proves especially valuable for theoretical investigations and grasping the optimal performance of quantum algorithms.

\subsection{Data and Metrics: \label{data}}

We evaluated the efficacy of our quantum machine learning model using the credit card fraud detection and breast cancer diagnostic datasets. These datasets serve as standard benchmarks for testing and comparing different machine learning models. Details about these models can be found in the Methods section.

\subsection{\textbf{Results on Credit Card Dataset:} \label{cc}}

\begin{figure}[ht]
\centering
\includegraphics[width=\linewidth]{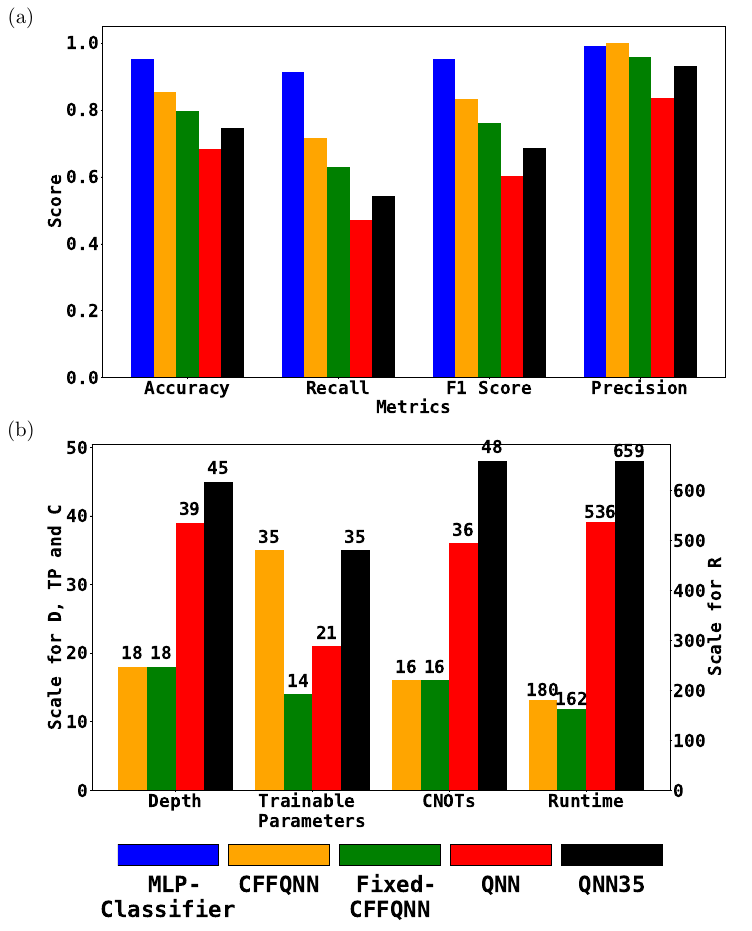}
\caption{(a) Performance comparison of various models analyzed on the credit card fraud detection dataset. (b) Comparative analysis of resources utilized by QNN and CFFQNN on this dataset. The scales for the depth, trainable parameters, and C-NOTs using the vertical axis on the left while that for the runtime uses the axis on the right. The bars from left to right are for the MLPClasifier, CFFQNN, FixedCFFQNN, QNN, and QNN35, omiting the bar for resources for the MLPClassifier in (b) because it is not a quantum model.}
\label{Fig6}
\end{figure}

For this study, we used the PCA to reduce the dimension of Credit card dataset to seven features. For both CFFQNN and FixedCFFQNN, we utilized a network structure with three nodes in the initial layer, two in the subsequent layer, and a single node in the concluding layer.

Figure \ref{Fig6}(a) presents a performance comparison between CFFQNN, FixedCFFQNN, QNN, and the classical MLPClassifier. The bar chart clearly demonstrates the superior performance of both CFFQNN variants over the conventional QNN model. Despite augmenting the QNN with 35 parameters for a balanced comparison, there was no notable enhancement in its results.

Figure \ref{Fig6}(b) outlines the quantum resources utilized by each quantum model. Evidently, the CFFQNN models require a reduced circuit depth, fewer CNot gates, and a shorter simulator runtime compared to the QNN model. Additionally, the FixedCFFQNN model exhibits a reduced requirement on the number of parameters.

\subsection{\textbf{Results on Breast Cancer Dataset:} \label{bc}}

\begin{figure}[ht]
\centering

\includegraphics[width=\linewidth]{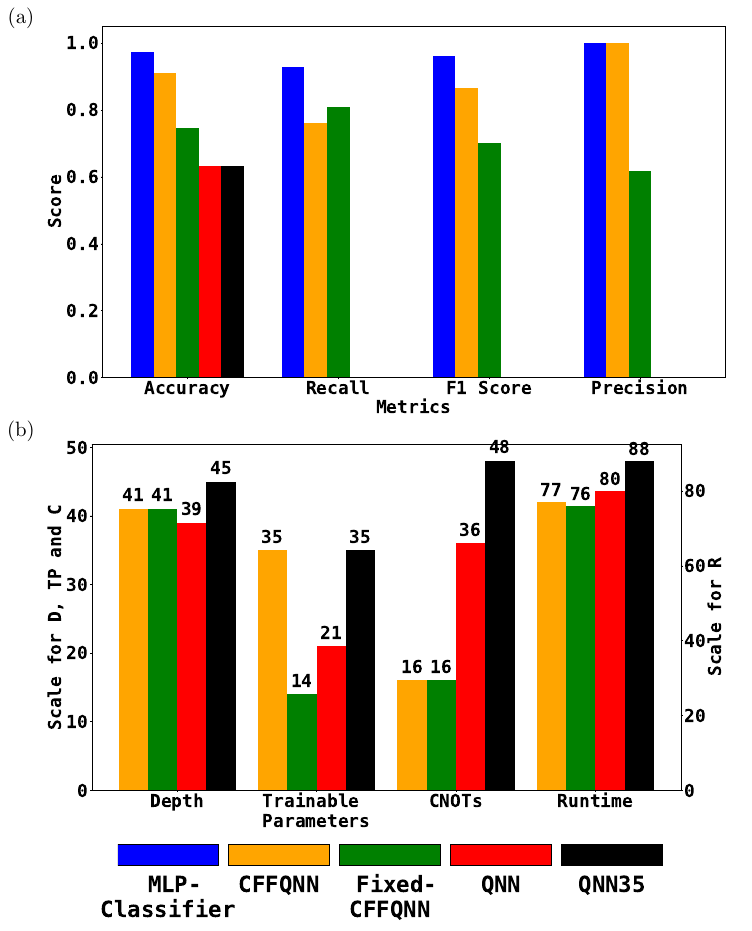}

\caption{(a) Performance comparison of various models analyzed on the breast cancer diagnostic dataset. (b) Comparative analysis of resources utilized by QNN and CFFQNN on this dataset. Scale and legend are the same as in Fig.~\ref{Fig6}.}
\label{Fig7}
\end{figure} 

For the breast cancer dataset, we narrowed down the feature space to 7 features and executed 100 iterations of COBYLA to enhance the quantum models. We adopted a network structure similar to that used for the credit card dataset in our proposed quantum models.

In Figure \ref{Fig7}(a), it is evident that both CFFQNN and FixedCFFQNN outperform the existing QNN models. The QNN's Recall, F1, and Precision scores are all zero, indicating a failure of the model to effectively learn from the data, resulting in the misclassification of all test data into a single category.

Figure \ref{Fig7}(b) provides a comparative analysis of the resources consumed by each of the four quantum models. While the resource utilization is largely consistent across all models, it is noteworthy that the QNN models demand a significantly higher number of C-NOT gates compared to the CFFQNN variants.  

\section{Discussion and Conclusion}

Our approach and results demonstrates a step forward in quantum machine learning through the introduction of the CFFQNN. 
Our CFFQNN is the direct quantum upgrade of a classical ANN, with all possible outputs of each classical perceptron upgraded to a branch of the quantum state that exists simultaneously and interacts with all other branches.
The advantages of the CFFQNN over previous QNNs stems from its unique integration of the 
$\sum_i X_{i}W_{i}+b$ term, inspired by classical neural networks, into the initial layer within a higher-dimensional Hilbert space. This approach, combined with the strategic incorporation of quantum entanglements in subsequent layers, marks a significant advancement over traditional QNN models. 

A pivotal advantage of our model is that its qubit count and number of C-NOT gates in its circuit are independent from the number of features in the dataset. This independence is particularly advantageous in dealing with complex, realistic datasets with varying feature sizes, ensuring scalability and flexibility. Furthermore, the model's design allows for direct adjustments in intermediate layers, a feature that enables it to effectively support deep network architectures, a limitation in many existing QNN models.

Through numerical experimentation, we have demonstrated the superior performance of CFFQNN in classification tasks, including a notable variant: the FixedCFFQNN. In the latter approach, we did not train any parameters in the first layer, yet it still outperformed existing QNN models. This untrained variant underscores the inherent efficiency and robustness of the CFFQNN architecture. Achieving high accuracy while requiring minimal quantum resources, both the standard and Fixed CFFQNN variants are significant steps toward practical use of quantum machine learning. Compared to QNN models utilizing the ZZFeature Map, the CFFQNN exhibits not only enhanced performance but also a more efficient utilization of quantum resources. These characteristics of our model will persuade realization on various quantum computing hardware as initial steps toward a scalable implementation for practical application of quantum machine learning.

%\st{The implications of these findings are substantial for the field of Quantum Machine Learning. They highlight the potential of integrating classical neural network concepts into quantum computing frameworks, opening avenues for more sophisticated, resource-efficient quantum models in the future. As we continue to explore and refine CFFQNN, we anticipate its adaptability to a broader range of applications, further bridging the gap between quantum computing and real-world machine learning challenges.}

\section{Methods}
We create an instance of the CFFQNN to perform data classification on two standard datasets. The CFFQNN model is simulated using Qiskit, while the datasets employed are accessible through Kaggle. We here detail parameters and techniques used in creating, training, and evaluating the model.

First, we select the maximum number of qubits to be used in our network as seven: this is small enough to be simulable on a standard personal computer yet large enough to exhibit genuinely quantum features such as a large Hilbert space spanned by 128 elements. The major question is whether this is a sufficient number of qubits to perform nontrivial machine learning tasks. Standard QNNs require one qubit per feature in the dataset, so we limit our investigation to data with $7$ features. In comparison, the CFFQNN can handle more features with the same total number of qubits, so we note that the number of features is limited by the QNNs against which we seek to compare the CFFQNN, not by the CFFQNN itself.

Next, we take real-world datasets and reduce their feature spaces to make the data size suitable for simulation, noting that many-qubit quantum systems are inherently challenging to simulate and that therein lies potential quantum advantages.
The Credit Card Fraud detection dataset has 30 features with which one seeks to evaluate whether a given transaction was fraudulent or not, a binary classification problem, while the Breast Cancer Diagnostic dataset's 30 features are used to evaluate whether a given patient does or does not have breast cancer, another binary classification. Since some of the features may be highly correlated with each other or may contribute less to the overall variance in the data distribution, a linear transformation of the coordinates in the feature space can elicit the principal components, which are the new coordinate axes that account for most of the independent information contained in the features and allow one to neglect axes where the data change less. Such principal component analysis (PCA) is standard in data processing and here reduces both 30-feature datasets to seven principal features each. These details are summarized in Table~\ref{tab:datasets methods}.

\begin{table*}[!hbt]
\centering
\caption{Properties of the datasets used to evaluate the CFFQNN and compare it to existing neural networks.}
\label{tab:datasets methods}
\begin{tabular}{|c|c|c|c|c|c|}
\hline
Datasets &
  Features &
  Features used &
  Training size &
  Testing size &
  Labels \\ \hline
\begin{tabular}[c]{@{}c@{}}\href{https://www.kaggle.com/datasets/mlg-ulb/creditcardfraud}{Credit card fraud detection (balanced)} \end{tabular} &
  30 &
  7 &
  688 &
  296 &
  2 \\ \hline
\href{https://archive.ics.uci.edu/dataset/17/breast+cancer+wisconsin+diagnostic}{Breast cancer diagnostic (Wisconsin)} &
  30 &
  7 &
  455 &
  114 &
  2 \\ \hline
\end{tabular}
\end{table*}

In the case of the Credit Card dataset, we also addressed the issue of class imbalance. To ensure unbiased training and evaluation, we eliminated the excess class instances, balancing the dataset. This step enabled our model to learn from both the minority and majority classes more effectively, thereby enhancing its ability to detect fraudulent transactions accurately.
By employing these preprocessing techniques on the selected datasets, we aimed to create a robust and reliable framework for evaluating the effectiveness of our quantum machine learning model.

The standard QNN is programmed as follows. Every qubit is initialized in the superposition state $(|0\rangle+|1\rangle)/\sqrt{2}$ by means of a Hadamard transformation, then the data features are encoded using a phase gate
\begin{equation}
    P(\Bar{X}_i)=|0\rangle\langle 0|+\eu^{\iu \bar{X}_i}|1\rangle\langle 1|
\end{equation} acting on the $i$th qubit; this is the ZFeatureMap with $\Bar{X}_i=2X_i$ and is the first stage of the ZZFeatureMap.  
The ZZFeatureMap then continues to sequentially entangle the $i$the and $j$th qubits and again upload the same data onto the quantum state, using the sequence of gates:

\begin{equation}
    G_{ij}=\mathrm{CNOT}^{i\mapsto j}[\mathds{I}_i\otimes P_j(X_{ij})]\mathrm{CNOT}^{i\mapsto j}
\end{equation}

which uses the controlled-not gate $\mathrm{CNOT}^{i\mapsto j}=(|0\rangle\langle 0|\otimes\mathds{I}+|1\rangle\langle 1|\otimes \sigma_x)$ and the nonlinear function of the parameters $X_{ij}=2(\pi-X_i)(\pi-X_j)$. All of the qubits are pairwise entangled using a sequence of $G_{ij}$ operators for various $i$ and $j$. However, the relationship between the number of C-NOT gates and the number of qubits is not fixed in a strict mathematical sense, but rather it depends on the specific architecture of the ZZFeatureMAp quantum circuit and the requirements of the algorithm being implemented. Here, we used the circuit with \textit{full} entanglement option, which requires $\frac{N(N-1)}{2}$ C-NOT gates to fully entangle all pairs of qubits for single repetition of the circuit. 

After all of the features are uploaded, the next step of the QNN is a parametrized quantum circuit. Single repetition of this consists of at least $2N$ single-qubit rotation gates $R_y(\theta_i)$, $N$ acting on each qubit, separated by fixed entangling gates. Each qubit experiences one parametrized $R_y$ gate, then a sequence of entangling gates $\prod_{i=1}^{i-1} \mathrm{CNOT}^{N-1\mapsto N}\cdots\mathrm{CNOT}^{2\mapsto 3}\mathrm{CNOT}^{1\mapsto 2}$ is applied, then the process is repeated in alternating fashion and ends with parametrized single-qubit rotation gates for a total of $2(N-1)$ controlled operations in the parametrized circuit. All the qubits are then measured in the computational basis and the measurement result is processed in the same was as for the CFFQNN detailed below. 

In comparison, the data may be encoded into any number of qubits for the CFFQNN, with more qubits being required for subsequent manipulations that correspond to hidden layers of ANNs. Just like in classical machine learning, there is no \textit{a priori} method for determining how many layers and how many nodes in each layer will be required for the success of training the network for a particular dataset. We choose to encode our datasets' seven features into three qubits, corresponding to the first hidden layer, process them with a second hidden layer comprised by two qubits, then funnel the quantum information into a final qubit such that the total number of qubits is only six.

The three qubits have the same data redundantly uploaded into them using no entangling operations: the operator $R_y(\sum_{i=1}^N W_i X_i+b)$ is applied to each of the first three qubits as in the main text. To process the data and forward it to the second hidden layer, a controlled operation is required between each pair of qubits from the first and second layers, such that twelve operations of the form $CR_y^{i\mapsto j}(\theta_{ij})$ with unique parameterized weights $\theta_{ij}$ are applied. A single-qubit rotation corresponding to a biasing term is also applied to each qubit in the second layer. Finally, three controlled operations $CR_y^{j\mapsto N}(\theta_{jN})$ are performed between the qubits in the second hidden layer and the final qubit along with a biasing term on the final qubit, for a total of 16 controlled operations. The final qubit is measured in the computational basis.

For both setups, the single output parameter $\langle Z_N\rangle$ is fed into a classical non-linear function and used to classify the input data. At least $70\%$ of the available data points are used and the models are scored on how well they correctly predict the classification of those data. The parameterized circuits are then updated with new parameters obtained from the COBYLA \cite{scikit-learn} optimizer and this process is repeated iteratively until the parameters converge. Those parameters are then used to test the models' performances on the test data points that they have not seen before. The overall models are then evaluated using the numbers of true positive (TP), false positive (FP), true negative (TN), and false negative (FN) results. These can be combined into a number of metrics as in Table~\ref{tab:metrics methods}. All of these metrics range from zero to one, with larger numbers implying superior models.
% \begin{table}[!hbt]
%     \centering
%     \begin{tabular}{ | m{5em} | m{4cm} | } 
%         \hline
%         Metrics & Equations \\
%         \hline
%         \hline
%         Precision & $\frac{\mathrm{TP}}{\mathrm{TP}+\mathrm{FP}}$ \\
%         \hline
%         Recall & $\frac{\mathrm{TP}}{\mathrm{TP}+\mathrm{FN}}$ \\
%         \hline
%         Accuracy & $\frac{\mathrm{TP}+\mathrm{TN}}{\mathrm{TP}+\mathrm{FP}+\mathrm{FN}+\mathrm{TN}}$ \\
%         \hline
%         F1-Score & $\frac{2\times \text{Precision} \times \text{Recall} }{\text{Precision} + \text{Recall}}$ \\
%         \hline
%     \end{tabular}
%     \caption{Metrics used to evaluate and compare the performances of different neural networks on the same datasets.}
%     \label{tab:metrics methods}
% \end{table}
\begin{table}[]
\centering
\caption{Metrics used to evaluate and compare the performances of different neural networks on the same datasets.}
\label{tab:metrics methods}
\begin{tabular}{|c|c|}
\hline
Metrics   & Equations                                     \\ \hline
Precision & $\frac{\mathrm{TP}}{\mathrm{TP}+\mathrm{FP}}$ \\ \hline
Recall    & $\frac{\mathrm{TP}}{\mathrm{TP}+\mathrm{FN}}$ \\ \hline
Accuracy & $\frac{\mathrm{TP}+\mathrm{TN}}{\mathrm{TP}+\mathrm{FP}+\mathrm{FN}+\mathrm{TN}}$         \\ \hline
F1-Score & $\frac{2\times \text{Precision} \times \text{Recall} }{\text{Precision} + \text{Recall}}$ \\ \hline
\end{tabular}
\end{table}

All of the source code for creating and comparing these models are detailed in the GitHub repository\_.

\section{Acknowledgments \label{secVI}}
The authors acknowledge that the NRC headquarters is located on the traditional unceded territory of the Algonquin Anishinaabe and Mohawk people. The authors would like to acknowledge the use of IBM Quantum services for this work and in particular the Qiskit package \cite{Qiskit}. AZG acknowledges support from NSERC's postdoctoral fellowship. KH acknowledges support from NSERC's Discovery Grant program.

% IBM, the IBM logo, and ibm.com are trademarks of International Business Machines Corp., registered in many jurisdictions worldwide. Other product and service names might be trademarks of IBM or other companies. A comprehensive list of IBM trademarks can be found at \href{https://www.ibm.com/legal/copytrade}{https://www.ibm.com/legal/copytrade}.
\section{Data Availability}
Data and code related to this research can be found at this private \href{https://github.com/utkarshh-singh/CFFQNN}{GitHub repository} upon reasonable request.

% \printbibliography 
%\bibliographystyle{apsrev4-2}
%\bibliography{refer}


\newpage
\clearpage

% \onecolumngrid
% \appendix
% \section*{Supplementary Material}

\end{document}